\documentclass[prl,10pt,floatfix,amsmath,twocolumn,showpacs,superscriptaddress]{revtex4}
\usepackage{graphicx}
\usepackage{dcolumn}
\usepackage{txfonts}
\usepackage{xcolor}
\usepackage{amssymb}
\usepackage{amsmath}
\usepackage{latexsym}
\usepackage{epsfig}
\usepackage{amsbsy}
\usepackage{array}
\usepackage{tabularx}

\begin{document}
\title{Combined Fast Reversible Liquid-like Elastic Deformation
with Topological Phase Transition
in Na$_3$Bi}

\author{Xiyue Cheng}
\affiliation{Shenyang National Laboratory for Materials Science,
Institute of Metal Research, Chinese Academy of Science, 110016
Shenyang, Liaoning, China}

\author{Ronghan Li}
\affiliation{Shenyang National Laboratory for Materials Science,
Institute of Metal Research, Chinese Academy of Science, 110016
Shenyang, Liaoning, China}

\author{Dianzhong Li}
\affiliation{Shenyang National Laboratory for Materials Science,
Institute of Metal Research, Chinese Academy of Science, 110016
Shenyang, Liaoning, China}

\author{Yiyi Li}
\affiliation{Shenyang National Laboratory for Materials Science,
Institute of Metal Research, Chinese Academy of Science, 110016
Shenyang, Liaoning, China}

\author{Xing-Qiu Chen}
\email[Corresponding author:]{xingqiu.chen@imr.ac.cn}
\affiliation{Shenyang National Laboratory for Materials Science,
Institute of Metal Research, Chinese Academy of Science, 110016
Shenyang, Liaoning, China}

\date{\today}

\begin{abstract}
By means of first-principles calculations, we identified the
structural phase transition of Na$_3$Bi from hexagonal ground state
to cubic $cF$16 phase above 0.8 GPa, in agreement with the
experimental findings. Upon the releasing of pressure, \emph{cF}16
phase of Na$_3$Bi is mechanically stable at ambient condition. The
calculations revealed that the $cF$16 phase is topological
semimetal, in similarity to well-known HgTe and it even exhibits an
unusually low $C^\prime$ modulus (only about 1.9 GPa) and a huge
anisotropy, $A^u$ of as high as 11, the third highest value among
all known cubic crystals in their elastic behaviors. These facts
render \emph{cF}16-type Na$_3$Bi very soft with a liquid-like
elastic deformation in the (110)$<$1$\overline{1}$0$>$ slip system.
Importantly, as accompanied with this deformation, Na$_3$Bi shows a
topological phase transition from a topological semimetal state at
its strain-free cubic phase to a topological insulating state at its
distorted phase. Because the $C^\prime$ elastic deformation almost
costs no energy in a reversible and liquid-like soft manner,
\emph{cF}16-type Na$_3$Bi would potentially provide a fast on/off
switching way between topological insulator and topological
semimetal, which would be beneficial to the quantum electronic
devices for practical applications.
\end{abstract}


\maketitle

The topological material have been extensively studied and its
number and types have increased dramatically over the past decades.
As the first member of the topological family, the topological
insulators (TIs) are materials with a bulk band gap but have
protected metallic states on their edge/surface, originating from
the combination of spin-orbit interactions and time-reversal
symmetry\cite{2010-Hasan, 2011-Qi, 2012-Yan}, including
Bi$_2$Se$_3$\cite{2009-Zhang,2009-Xia} and the topological
crystalline insulators (TCIs) such as
Pb$_{1-x}$Sn$_x$Te(Se), SnSe, and SnS\cite{2012-Xu,2012-Fu, 2013-Yan}. In anglogy to TIs, the
non-trivial topology of the band structure in semimetal gives rise
to other new states of matters, topological Dirac semimetals (TDSs)
and topological Weyl semimetals (TWSs). For both TDSs and TWSs, the
bulk conduction and valence bands touch only at discrete (Dirac)
points and disperse linearly along all momentum directions, while
their low-energy bulk excitations are described by the Dirac and
Weyl equations, respectively. Compared to Dirac semimetal, inversion
or time reversal symmetry must be broken in Weyl semimetal. For
example, Na$_3$Bi\cite{2012-Wang,2014-Liu1, 2014-Cheng, 2015-Xu} and
Cd$_3$As$_2$ \cite{2013-Wang2, 2014-Ali, 2014-Liu2} were
theoretically and experimentally demonstrated to be 3D TDSs
protected by crystal symmetry and the noncentrosymmetric materials
including TaAs, TaP, NbAs, and NbP \cite{2015-Wang} was recently
predicted as natural TWSs. Certainly, the topological concept can be
also introduced into metals leading to topological metals (TMs).
Generally, most of TMs can be manually achieved from TIs or
semimetals, yet so far several compounds, such as HgTe
\cite{2007-Konig,2006-Bernevig2} and NaBi \cite{2015-Li}, were found
to be a native 3D TM. Although the flourishing topological material
has inspired great interest, one has to admit the fact that some
existed challenges are restricting its development, no matter from
the aspects of scientific researches or technological application.

Besides two often facing challenges (small band gap and small
quantized conductance of the edge state)
\cite{2009-Zhang,2009-Xia,2007-Konig,2014-Qian}, another main
challenge is the lacking of effective ways \cite{2014-Qian} to
control the phase transition between topological materials and
regular insulators or metals. Traditionally, the control of
topological phase transitions was realized mainly by pressure,
lattice strain, or chemical substitution. For instance, $hP$8-type
Na$_3$Bi undergoes a topological phase transition from a Dirac
semimental to a trivial insulator upon doping Sb or P through
varying the concentrations of Na$_3$Bi$_{1-x}$(Sb,P)$_x$
\cite{Na3Bi-2015}. However, via such a way it is extremely difficult
to make the transition reversible once the material is fabricated.
But, is there any way to rapidly and effectively switch on or off
these phase transitions, rather than traditional means? This
definitely poses a challenging issue.

\begin{figure*}[hbt]
\centering
\includegraphics[width=0.95\textwidth]{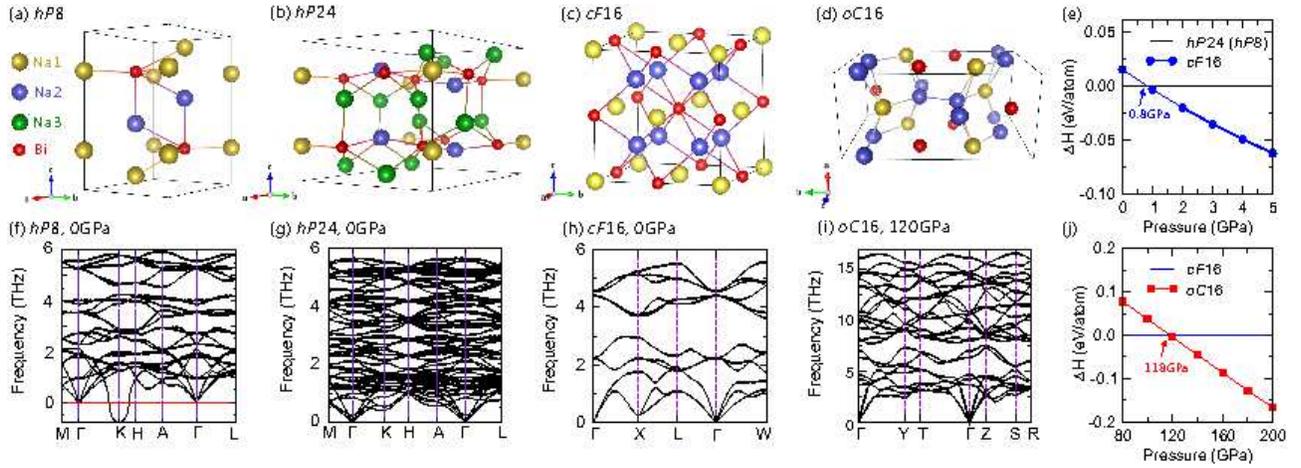}
\caption{(Color online) The DFT-derived pressure-dependent phase
transitions of Na$_3$Bi. (a) $hP$8 hexagonal phase, which was
experimentally claimed to be stable at the ground state under
ambient pressure, (b) $hP$24 hexagonal phase, which was
theoretically found to be stable at the ground state under ambient
pressure, (c) $cF$16 cubic phase, which is more stable in energy
above 0.8 GPa than both $hP$8 and $hP$24 phases, and (d) $oC$16
orthorhombic phase, which is more stable above 118 GPa. (f,g,h,i)
the derived phonon dispersions of $hP$8 (0 GPa), $hP$24 (0 GPa),
$cF$16 (0 GPa) and $oC$16 (120 GPa), respectively. (e and j) the
pressure-dependent enthalpies of $hP$8, $hP$24, $cF$16 and $oC$16
phases to show the pressures of phase transitions. The red circles
represent the Bi atoms while the Na atoms are shown in different
color according to their Wyckoff sites. Here, yellow, blue and green
circles represents Na1, Na2 and Na3 atoms, respectively. }
\label{fig1}
\end{figure*}

As advanced functional materials, the elastic or mechanical
properties of topological materials were often neglected and
concealed under their wonderful and shining electronic properties.
It is true that a vast majority of topological materials only have
ordinary elastic or mechanical properties which may just indicate
the structural mechanical stability \cite{2014-Sa}. Thus, it would
be highly interesting to find a topological material with extinct
elastic or mechanical properties. Recently, Na$_3$Bi was proposed
and demonstrated to be a 3D TDS \cite{2012-Wang,2014-Liu1,
2014-Cheng, 2015-Xu}, and it is found that Na$_3$Bi would undergo a
structural phase transition of from its hexagonal ground state
(\emph{hP}24) to a cubic $cF$16 phase at 0.8 GPa \cite{2015-Cheng},
in well agreement with the experimental findings \cite{1998-Leonova,
2003-Leonova, 1999-Kulinich}. To one's surprise, this
$cF$16-Na$_3$Bi exhibits an unusually low $C^\prime$ modulus (only
about 1.9 GPa) and a huge anisotropy, $A^u$ of as high as 11, the
third highest value among all known cubic crystals in its elastic
properties. These facts render \emph{cF}16-type Na$_3$Bi very soft
with a liquid-like elastic deformation in the
(110)$<$1$\overline{1}$0$>$ slip system. Meanwhile, as the
$C^\prime$ elastic deformation induced a topological phase
transition from TS to TI and it almost cost no energy in a
reversible and liquid-like elastic manner, this cubic Na$_3$Bi phase
would potentially provide a fast on/off switching way between TI and
TS.

The structural optimization, electronic properties and elastic
behaviors of the Na$_3$Bi phases were calculated within the
framework of density functional theory ({\small DFT})
\cite{Hohenberg,Kohn} using the Vienna \emph{ab initio} Simulation
Package ({\small VASP}) \cite{Kresse-1, Kresse-2} with the projector
augmented wave ({\small PAW}) method \cite{PAW} and generalized
gradient approximation ({\small GGA}) within the
Perdew-Burke-Ernzerhof ({\small PBE}) exchange-correlation
functional \cite{PBE}. The cutoff energy for the expansion of the
wavefunction into plane waves was set at 350 eV and the tetrahedron
method with Bl\"{o}chl corrections was ultilized. All the Brillouin
zone integrations were performed on Monkhorst-Pack $k$-meshes and
was sampled with a resolution of 2 $\pi$ $\times$ 0.07 \AA\ $^{-1}$,
which showed excellent convergence of the energy differences and
stress tensors. For hexagonal structures, the Brillouin zone
integrations were performed on the $\Gamma$-centered symmetry. To
check the dynamical stability, we further derived the phonon
dispersion curves using the finite-displacement approach as
implemented in the \emph{Phonopy} code \cite{phonopy}. The phonon
frequencies are constructed from forces, resulting from
displacements of certain atoms in a supercell containing typically
80-100 atoms for each Na$_3$Bi phases, respectively. The electronic
localized function (ELF)\cite{ELF, ELF1, ELF2} was done using the
grid-based algorithm with 100 $\times$ 100 $\times$ 100 grids. In
addition, all crystal structures and ELF diagrams were generated
using {\small VESTA} \cite{VESTA}.

The latest discoveries that native Na$_3$Bi is a three dimensional
(3D) Dirac semimetal represent a significant advance in topological
Dirac materials \cite{2012-Wang, 2014-Cheng, 2014-Liu1}, because
this material enables the study of a new type of quantum state. At
ambient condition, Na$_3$Bi possesses bulk Dirac fermions in 3D at
the Fermi level, which disperse linearly along all three momentum
directions \cite{2012-Wang,2014-Cheng}, in contrast to the
two-dimensional Dirac fermions present on the surfaces of 3D
topological insulators and in graphene \cite{2009-Castro}.

Recently, the first-principles calculations revealed that the
previously experimentally characterized $hP$8 phase (Fig.
\ref{fig1}a) is unstable at the ground state due to the presence of
a negative phonon branch around the high-symmetry X point, as
illustrated in Fig. \ref{fig1}(f). Instead, the calculation
suggested that the real ground-state phase would be the $hP$24 phase
(Fig. \ref{fig1}(b)), which is indeed a distorted superlattice
version of the $hP$8 phase \cite{2014-Cheng}. Importantly, this
stable $hP$24 structure exhibits a semblable 3D TDS feature as what
the $hP$8 phase presents \cite{2012-Wang}.

Furthermore, the previous experiments demonstrated that Na$_3$Bi
undergoes a phase transition above the pressures of 0.7-1.0
GPa\cite{1998-Leonova, 2003-Leonova, 1999-Kulinich}. In nice
agreement with the experimental measurements, our first-principles
calculations successfully reproduced this phase transition from the
ground-state $hP$24 (or $hP$8) phase to the cubic $cF$16 phase (Fig.
\ref{fig1}(c)) above 0.8 GPa \cite{2015-Cheng}, as shown in Fig.
\ref{fig1}(e). The calculations demonstrated that, even upon the
releasing pressure, the $cF$16 phase is still mechanically and
dynamically stable at zero pressure, as elucidated by its phonon
dispersions (Fig. \ref{fig1}(h)). This fact indicates that the
$cF$16-type phase is quenchable at ambient condition once it is
synthesized above 0.8 GPa. Specifically, this cubic Na$_3$Bi
crystalizes in a BiF$_3$-type structure (space group of
$Fm$$\overline{3}$$m$) with $a$= 7.550 \AA\,at ambient zero
pressure. The Bi atoms lie at 4$b$ site (0.5, 0.5, 0.5) and the Na
atoms occupy two inequivalent sites, 4$a$ (0, 0, 0) and 8$c$ (3/4,
3/4, 3/4). The calculation also revealed that, by further increasing
the pressure up to above 118 GPa, the cubic $cF$16 phase can be
transformed into an orthorhombic $oC$16 phase (Fig. \ref{fig1}(d)
and Fig. \ref{fig1}(j)), which is a wide-gap insulator
\cite{2015-Cheng}.

\begin{figure}[hbt]
\centering
\includegraphics[width=0.47\textwidth]{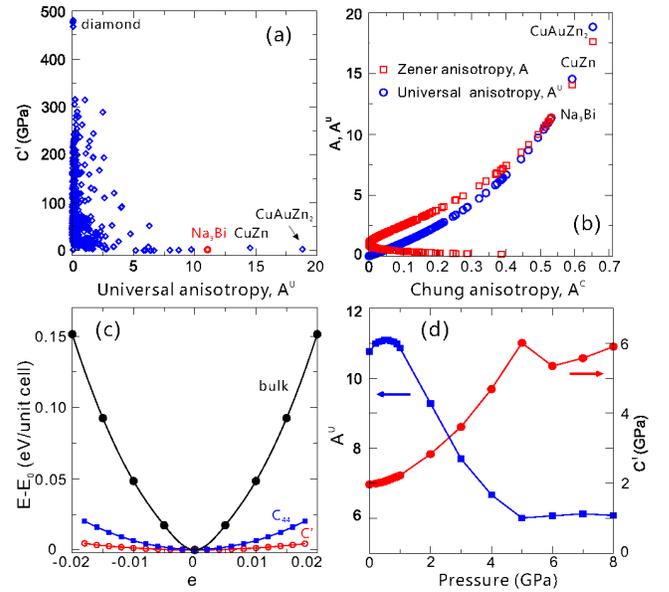}
\caption{(Color online) The liquid-like elastic behavior of
Na$_3$Bi. (a) $C^\prime$ versus universal anisotropic ratio $A^u$,
showing that Na$_3$Bi has an extremely low $C^\prime$ and a highly
large $A^u$ among a variety of cubic crystals, (b) anisotropic ratio
$A$ ($A^u$) versus Chung  anisotropy A$^c$ for a variety of cubic
crystals, (c) the DFT-derived deformation energy as a functional of
strain according to the definition of elastic energies of bulk
modulus, C$_{44}$, and C$^\prime$ at 0 GPa, (d) the
pressure-dependent universal anisotropic ratio $A^u$ and
$C^\prime$.} \label{fig2}
\end{figure}

Surprisingly, this cubic $cF$16 phase exhibits several unusual
elastic behaviors in its mechanical properties. At 0 GPa, the
DFT-derived single crystal elastic constants are $C_{11}$ = 22.3
GPa, $C_{12}$ = 18.4 GPa and $C_{44}$ = 21.9 GPa. Firstly, C$_{11}$
is very close to C$_{44}$ , which gives nearly coinciding
longitudinal and transverse acoustic phonons along the direction of
$<$001$>$. This fact is reflected well by that two phonon branches
along this $\Gamma$-X direction (Fig. \ref{fig1}(h)) are nearly
degenerated at small $q$ values close to the $\Gamma$ point.
Secondly, it has been noted that the difference of $C_{44}$-$C_{12}$
is only about 3.5 GPa, revealing a very small deviation from the
Cauchy relation with C$_{44}$ = C$_{12}$, which measures the
importance of the angular dependence in atomic forces as compared to
the central force description. Thirdly, $cF$16-Na$_3$Bi has an
rather low $C^\prime$ = $\frac{C_{11}-C_{12}}{2}$ = 1.9 GPa. Note
that $C^\prime$ represents the resistance to shear deformation by a
shear stress applied across the (110) plane in the
$<$1$\overline{1}$0$>$ direction and it indeed measures the rigidity
against the volume-conserving tetragonal deformation. We further
derived its universal anisotropic ratio to be as high as $A^u$ =
11.07, according to the recently proposed definition of the
universal elastic anisotropy index
 \cite{2008-Ranganathan}, highlighting a huge anisotropy
in its elastic properties. Furthermore, we plot the $C^\prime$
versus $A^u$ for a variety of cubic crystals (330 compounds, elastic
data from the supplementary table of ref. \cite{2012-Niu}) as
illustrated in Fig. \ref{fig2}(a and b). Among those compounds, we
have observed from the plot that the cubic Na$_3$Bi almost exhibits
the lowest $C^\prime$ value and the third highest anisotropic ratio
$A^u$ (which are only lower than two known superelastic compounds of
CuZn and CuAuZn$_2$)\cite{CuZn,CuZnAu2}. These unique features
render $cF$16-Na$_3$Bi special: it is very soft, showing a
liquid-like behaviour along the $C^\prime$ deformation corresponding
to the shear slip (110)$<$1$\overline{1}$0$>$ system within the
elastic regime. This fact has also been evidenced by Fig.
\ref{fig2}(c), from which it can be seen that the $C^\prime$
deformation almost costs no energy, based on the deformation
energy-versus-strain relation. In particular, it has been also found
that, the $C^\prime$ and universal anisotropic ratio $A^u$ show
unusual trends with increasing pressure, as evidenced in Fig.
\ref{fig2}(d). In the first range below 1 GPa, both $C^\prime$ and
$A^u$ remain almost constant and, in the pressure range from 1 GPa
to 4 GPa $C^\prime$ rapidly increases from 2 GPa to 6 GPa whereas
$A^u$ shows an apparent drop from about 11 to 6. Above about 4 GPa,
they again almost become constant. In particular, the liquid-like
and soft elastic behavior of Na$_3$Bi remains robust in a certain
pressure range at least from 0 GPa to 1 GPa.

\begin{figure}[hbt]
\centering
\includegraphics[width=0.48\textwidth]{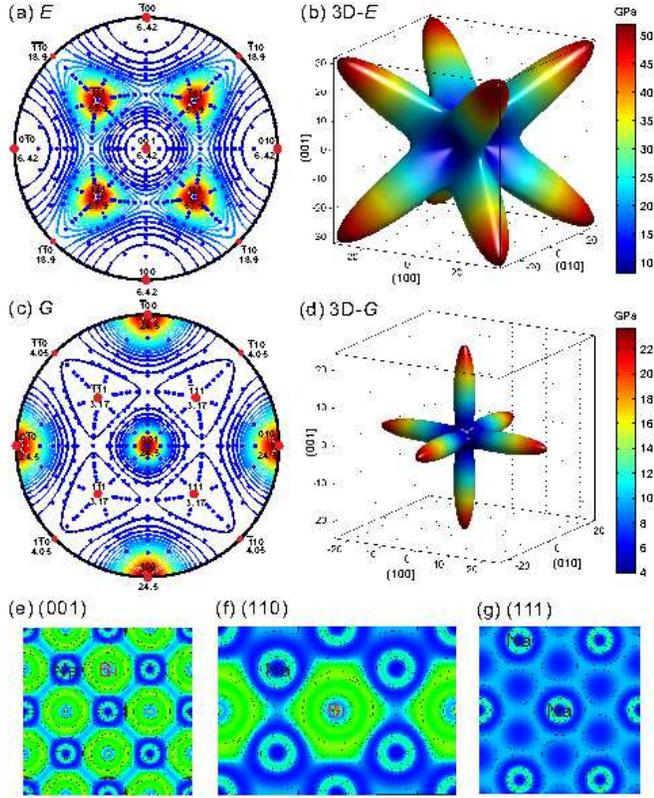}
\caption{(Color online) Huge elastic anisotropy of Na$_3$Bi. (a and
c) The stereographic projections of the Young's and shear moduli,
respectively. (b and d) The representation surfaces for the
crystallographic-orientation depended Young's and shear moduli,
respectively. (e, f, and g) The electron localization function (ELF)
isosurface maps for the (001), (110) and (111) planes, respectively.
} \label{fig3}
\end{figure}

Importantly, the huge anisotropic ratio of \emph{cF}16-type Na$_3$Bi
has been further reflected by the stereographic projections at
ambient pressure in Fig. \ref{fig3}(a and c), which are used in
order to show the particular crystal orientation and the
corresponding Young's and shear moduli more clearly, while the 3D
representation surfaces (Fig. \ref{fig3}(b and d)) visualized the
crystallographic-orientation dependent Young's and shear moduli
which, however, presents the strong and weak directions in one
crystal structure directly\cite{2014-Cheng2}. Note that the negative
sign only denotes the negative direction corresponding to the
positive one. In its stereographic projections, each blue point
denotes one crystal orientation and the lines with different colour
and density represent the counter lines of Young's or shear moduli.
It can be seen that the elastic properties under different crystal
orientations highly vary from each other. For Young's modulus, the
lowest and highest orientations are [001] ($E_{[001]}$ = 6.42 GPa)
and [111] ($E_{[111]}$ = 53.9 GPa), respectively. In contrast, its
shear modulus exhibits an opposite trend. The [001] orientation has
the highest shear modulus of $G_{[001]}$ = 24.5 GPa and the [111]
one with the lowest value of $G_{[111]}$ =3.17 GPa. There is no
doubt that their opposite trends between both [001] and [111]
directions substantially contribute to the huge anisotropy of this
cubic crystal. Mechanically, the huge anisotropy can be attributed
to the electronic structures and atomic arrangements. As illustrated
in the electronic localization function (ELF) images (Fig.
\ref{fig3}(e,f,g)), the (001) plane consists of Bi and Na atoms
within the ionic-bonding manner (Fig. \ref{fig3}(e)), whereas the
(111) plane is indeed comprised by the densest metallic Na atoms
with pure metallic bonds (Fig. \ref{fig3}(g)). It is clear that the
dense (111) plane is hard to be compressed, but it can be easily
sheared under deformation due to the pure metallic bonds. However,
for the (001) plane, all bonds are ionic Na-Bi bonds with low atomic
density. This interprets as to why the resistance to the compression
of the (001) plane is low, whereas its shear modulus is much higher.
Similar interpretation also holds for the (110) plane (Fig.
\ref{fig3}(f)),  from which we can see that both Na-Na metallic
bonds and Na-Bi ionic bonds co-exist (\emph{c.f.} and the metallic
bonds form a pure channel along the $<110>$ direction, Fig.
\ref{fig3}f), leading to the intermediate Young's and shear moduli.

\begin{figure}[hbt]
\centering
\includegraphics[width=0.48\textwidth]{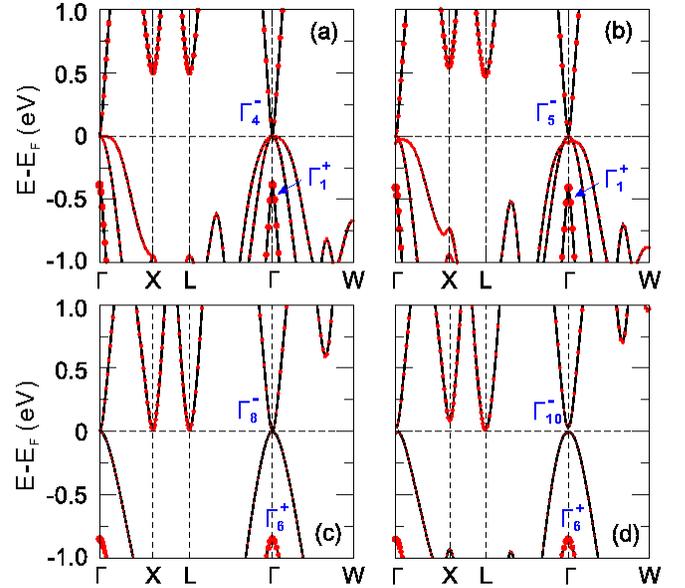}
\caption{(Color online) Topological phase transition from
topological semimetal to topological insulator under the $C^\prime$
tetragonal deformation. Band structures for cubic $cF$16-type
Na$_3$Bi (a and c) and for its distorted structure via the
$C^\prime$ deformation with a 4\% strain (b and d). a and b: no SOC
inclusion; c and d: with SOC inclusion. The solid circles denote the
$s$-like states.} \label{fig4}
\end{figure}

In addition to its unusual elastic properties, the cubic $cF$16
phase of Na$_3$Bi exhibits another type of attractive electronic
properties, in different from the ground-state \emph{hP}24 phase
with unique feature of 3D TDS. From Fig. \ref{fig4}(a), it has been
seen that the Fermi level exactly crosses the threefold degenerated
Bi-$p_{(x,y,z)}$ states ($\Gamma_4$) at $\Gamma$ for $cF$16-type
Na$_3$Bi at 0 GPa. If the spin-orbit coupling (SOC) effect is
included, this threefold degenerated $\Gamma_4$ \emph{p}-states
would further split into the doubly degenerated
$|$$p_{\frac{3}{2}}$$\rangle$ and nondegenerated
$|$$p_{\frac{1}{2}}$$\rangle$ states, as shown in Fig.
\ref{fig4}(c). With this electronic feature, it is clear that in the
strain-free cubic case Na$_3$Bi is a semimetal because its
conduction and valence bands touch only at the $\Gamma$ point.
Another typical feature is that even in the case without SOC, the
Na-3$s$ states ($\Gamma_1$) are energetically lower by about 0.4 eV
than the Bi-6$p$ states ($\Gamma_4$) at the $\Gamma$ point (Fig.
4(a)). More importantly, $\Gamma_1$ and $\Gamma_4$ exhibits the
opposite parities. However, it is worth noting that no any other
band inversions present for the rest high symmetric points in the
BZ. These facts imply that Na$_3$Bi naturally processes the inverted
band ordering, which is a prerequisite for the occurrence of
topological electronic properties. Interestingly, when the SOC is
switch on, the inverted band ordering is further enhanced. The
Na-3$s$ state ($\Gamma_6^+$ in Fig. \ref{fig4}(c)) is significantly
reduced in energy by increasing the energy difference up to 0.74 eV
(Fig. 4(c)). Therefore, in its strain-free cubic case,
$cF$16-Na$_3$Bi is a TS at 0 GPa, no matter whether the spin-orbit
coupling effect (SOC) is considered. Note that this topological
semimetal feature of $cF$16-Na$_3$Bi remains unchanged up to 3.65
GPa before it transforms to a regular insulator \cite{2015-Cheng}.
Certainly, it needs to be emphasized that the band inversion at
$\Gamma$ is caused by the crystal field effect via the protection of
lattice symmetry, rather than the chemical doping, pressure or
strains. This feature is similar to the famous case of HgTe with a
nearly zero direct band gap at $\Gamma$ point in the BZ and an
analogous inverted band ordering \cite{2006-Bernevig2}.

The doubly degenerated $|$$p_{\frac{3}{2}}$$\rangle$ states crossing
the Fermi level in Fig. \ref{fig4}(c) can be certainly broken by the
engineering of strains. Importantly, via the liquid-like $C^\prime$
soft deformation of the (110)$<$1$\overline{1}$0$>$ slip system, our
calculation demonstrated that the $cF$16-Na$_3$Bi can be indeed
transformed into a TI from its intrinsic feature of a TS. From Fig.
\ref{fig4}(b and d) the band structure at around the Fermi level
depends sensitively on the lattice distortion along the soft
(110)$<$1$\overline{1}$0$>$ tetragonal elastic deformation. Without
the SOC inclusion any slight lattice distortion along the $C^\prime$
deformation breaks up the triply degenerated Bi-$p_{(x,y,z)}$ states
($\Gamma_4$) at $\Gamma$ into doubly degenerated Bi-$p_{x,y}$ and
nondegenerated Bi-$p_z$ states (see Fig. \ref{fig4}(b) with a 4\%
distortion). Furthermore, with the SOC inclusion, the doubly
degenerated Bi-$p_{x,y}$ states at the Fermi level is further broken
to open a band gap, as evidenced in Fig. \ref{fig4}(d). In the
meanwhile, such a distortion does not affect its band inversion.
From Fig. \ref{fig4}(d), the nondegenerated Na-3$s$ state
($\Gamma_6^+$) is energetically lower by about 0.76 eV than the
Bi-$p$ state ($\Gamma_{10}^-$) at the $\Gamma$ point. It needs to be
emphasized that for cubic structure the product of the parities of
the Bl\"och wave function for the occupied bands at time-reversal
invariant momenta (TRIMs) of three equivalent X (0.5,0,0), three
equivalent M (0.5,0.5,0), and single R (0.5,0.5,0.5) has been
derived to be -1, whereas at $\Gamma$ it is +1. Via the $C^\prime$
deformation, although that these three equivalent momenta of X (or
M) can be lifted, the product of the parities for all occupied bands
at the TRIMs of $\Gamma$, R, X and M still remains -1. Therefore,
these results evidence that the distorted cubic Na$_3$Bi by the
strain engineering along the $C^\prime$ deformation is a
three-dimensional topological insulator. Importantly, because of the
reversible liquid-like elastic behavior along the $C^\prime$
deformation which almost cost no energy, the $cF$16-type Na$_3$Bi
becomes a highly attractive compound. Through such a way, one would
easily realize rapid on/off switching from a TS state in its cubic
phase to a TI state in its distorted lattice.

In summary, we have demonstrated that a soft liquid-like elastic
control of topological phase transition in cubic $cF$16-type
Na$_3$Bi can be achieved through the $C^\prime$ tetragonal
deformation along the (110)$<$1$\overline{1}$0$>$ slip system,
highlighting a realization of a fast and reversible on/off switching
way between topological semimetal and topological insulator. This
would be beneficial to the quantum electronic devices for practical
applications.

\bigskip
\noindent {\bf Acknowledgments} This work was supported by the
''Hundred Talents Project'' of the Chinese Academy of Sciences and
by the Key Research Program of Chinese Academy of Sciences (Grant
No. KGZD-EW-T06) and NSFC of China (Grand Numbers: 51074151). The
computational resource is using the local HPC cluster of the
Materials Process Modeling Division in the IMR as well as the
National Supercomputing Center in Tianjin (TH-1A system).


\begin{thebibliography}{18}
\bibitem{2010-Hasan}
M. Z. Hasan and C. L. Kane, Rev. Mod. Phys. \textbf{82}, 3045
(2010).

\bibitem{2011-Qi}
X.-L. Qi and S.-C. Zhang, Rev. Mod. Phys. \textbf{83}, 1057 (2011).

\bibitem{2012-Yan}
B. Yan and S. C. Zhang, Rep. Prog. Phys. \textbf{75}, 096501 (2012).

\bibitem{2009-Zhang}
H. Zhang, C.-X. Liu, X.-L. Qi, X. Dai, Z. Fang and S.-C. Zhang, Nat.
Phys., \textbf{5}, 438 (2009).

\bibitem{2009-Xia}
Y. Xia, D. Qian, D. Hsieh, L. Wray, A. Pal, H. Lin, A. Bansil, D.
Grauer, Y. S. Hor, R. J. Cava and M. Z. Hasan, Nat. Phys.
\textbf{5}, 398 (2009)


\bibitem{2012-Fu}  T. H. Hsieh, H. Lin,	J. Liu,	
W. Duan, A. Bansil and L. Fu, Nat. Commun. \textbf{3}, 982 (2013).

\bibitem{2012-Xu}
S. Y. Xu, C. Liu, N. Alidoust, M. Neupane, D. Qian, I. Belopolski, J. D. Denlinger, Y. J. Wang, H. Lin, L. A. Wray, G. Landolt, B. Slomski, J. H. Dil, A. Marcinkova, E. Morosan, Q. Gibson, R. Sankar, F. C. Chou, R. J. Cava, A. Bansil and M. Z. Hasan,
Nat. Commun. \textbf{3}, 1192 (2013).

\bibitem{2013-Yan} Y. Sun, Z.C. Zhong, T. Shirakawa, C. Franchini, D. Z. Li, Y.Y. Li, S. Yunoki, and X.-Q. Chen,
Phys. Rev. B \textbf{88}, 235122 (2013).

\bibitem{2012-Wang}
Z. Wang, Y. Sun, X.-Q. Chen, C. Franchini, G. Xu, H. Weng, X. Dai, and Z. Fang,
Phys. Rev. B \textbf{85}, 205101 (2012).

\bibitem{2014-Liu1}
Z. K. Liu, B. Zhou, Y. Zhang, Z. J. Wang, H. M. Weng, D. Prabhakaran, S.-K.  Mo, Z. X. Shen, Z. Fang, X. Dai, Z. Hussain, and Y. L. Chen,
Science \textbf{343}, 864 (2014).

\bibitem{2014-Cheng} X. Y. Cheng, R. H. Li, X.-Q. Chen, D. Z. Li, and Y. Y. Li,
Phys. Rev. B \textbf{89}, 245201 (2014).

\bibitem{2015-Xu} S. Y. Xu, C. Liu, S. K. Kushwaha, R. Sankar, J. W. Krizan, I. Belopolski, M. Neupane, G. Bian, N. Alidoust, T.-R. Chang, H.-T. Jeng, C.-Y. Huang, W.-F. Tsai, H. Lin, P. P. Shibayev, F.-C. Chou, R. J. Cava and M. Z. Hasan,
Science \textbf{347}, 294 (2015).

\bibitem{2013-Wang2}
Z. Wang, H. Weng, Q. Wu, X. Dai, and Z. Fang, Phys. Rev. B
\textbf{88}, 125427 (2013).

\bibitem{2014-Ali}
M. N. Ali, Q. Gibson, S. Jeon, B. B. Zhou, A. Yazdani and R. J.
Cava, Inorg. Chem. \textbf{53}, 4062 (2014).

\bibitem{2014-Liu2}
Z. K. Liu, J. Jiang, B. Zhou, Z. J. Wang, Y. Zhang, H. M. Weng, D.
Prabhakaran, S. K. Mo, H. Peng, P. Dudin, T. Kim, M. Hoesch, Z.
Fang, X. Dai, Z. X. Shen, D. L. Feng, Z. Hussain and Y. L. Chen,
Nat. Mater. \textbf{13}, 677 (2014).

\bibitem{2015-Wang}
H. Weng, C. Fang, Z. Fang, A. Bernevig and X. Dai,
Phys. Rev. X \textbf{5}, 011029 (2015).

\bibitem{2006-Bernevig2}
B. A. Bernevig, T. L. Hughes and S. C. Zhang,
Science \textbf{314}, 1757 (2006).

\bibitem{2007-Konig}
M. Konig, S. Wiedmann, C. Brune, A. Roth, H. Buhmann, L. Wiedmann,
X. L. Qi and S. C. Zhang, Science, \textbf{318}, 766 (2007).

\bibitem{2014-Qian} X. F. Qian, J. W. Liu, L. Fu and J. Li, Science,
\textbf{346}, 1344 (2014).

\bibitem{2015-Li}
R. H. Li, X. Y. Cheng, Q. Xie, Y. Sun, D. Z. Li, Y. Y. Li and X.-Q.
Chen, Sci. Rep., \textbf{5}, 8446 (2015).

\bibitem{Na3Bi-2015} A. Narayan, D. Disante, S. Picozzi, and S.
Sanvito, Phys. Rev. Lett., {\bf 113}, 256403 (2014).













\bibitem{2014-Sa}
B. Sa, J. Zhou, R. Ahuja and Z. Sun, Comput. Mater. Sci.
\textbf{82}, 66 (2014).

\bibitem{2015-Cheng}
X. Y. Cheng, R. H. Li, D. Z. Li, Y. Y. Li and X.-Q.Chen,
Phys. Chem. Chem. Phys. \textbf{17}, 6933 (2015).

\bibitem{1998-Leonova} M. E. Leonova, S. A. Kulinich,
and L. G. Sevast'yanova, Exp. Geosci. \textbf{7}, 55 (1998).

\bibitem{2003-Leonova} M.E. Leonova, I. K. Bdikin,
S.A. Kulinich, O.K. Gulish, L.G. Sevast'yanova, K.P. Burdina, Inorg.
Mater. \textbf{39}, 266 (2003).

\bibitem{1999-Kulinich} S. A. Kulinich, M. E. Leonova,
and L. G. Sevast'yanova, Zh. Obshch. Khim. \textbf{69}, 681 (1999).

\bibitem{Hohenberg} P. Hohenberg,
    Phys. Rev. B \textbf{136}, 864 (1964).

\bibitem{Kohn} W. Kohn and L. J. Sham,
    Phys. Rev. A \textbf{140}, 1133 (1965).

\bibitem{Kresse-1}  G. Kresse and J. Hafner,
    Phys. Rev. B \textbf{47}, 558 (1993).

\bibitem{Kresse-2}  G. Kresse and J. Furthm\"{u}ller,
    Phys. Rev. B \textbf{54}, 11169 (1996).

\bibitem{PAW}  P. E. Bl\"{o}chl,
    Phys. Rev. B \textbf{50}, 17953 (1994).

\bibitem{PBE}  J. P. Perdew, K. Burke, and M. Ernzerhof,
    Phys. Rev. Lett. \textbf{77}, 3865 (1996).

\bibitem{phonopy} A. Togo, F. Oba, I. Tanaka,
    Phys. Rev. B \textbf{78}, 134106 (2008).

\bibitem{ELF} A. D. Becke and K. E. Edgecombe,
J Chem. Phys. \textbf{92}, 5397 (1990).

\bibitem{ELF1} B. Silvi and A. Savin,
Nature \textbf{371}, 683 (1994).

\bibitem{ELF2} A. Savin, R. Nesper, S. Wengert,
and T. F. F\"assler, Angew. Chem. Int. Ed. Engl. \textbf{36}, 1808
(1977).

\bibitem{VESTA} K. Momma and F.  Izumi,
J. Appl. Crystallogr. \textbf{44}, 1272 (2011).

\bibitem{2009-Castro} A. H. Castro Neto, N. M. R. Peres,
K. S. Novoselov and A. K. Geim, Rev. Mod. Phys., \textbf{81}, 109 (2009).

\bibitem{2008-Ranganathan} S. I. Ranganathan and M. Ostojia-Starzewski,
Phys. Rev. Lett., {\bf 101}, 055504 (2008).

\bibitem{2012-Niu}
H. Niu, X.-Q. Chen, P. Liu, W. Xing, X. Cheng, D. Li and Y. Li, Sci.
Rep. \textbf{2}, 718 (2012).

\bibitem{CuZn} G. M. McManus, Phys. Rev. {\bf 129}, 2004 (1963).

\bibitem{CuZnAu2} M. Meyers and K. K. Chawla, in {\em Mechanical
Behavior of Materials}, Cambridge University Press, (2009).

\bibitem{2014-Cheng2}
X. Y. Cheng, X.-Q. Chen, D. Z. Li and Y. Y. Li, Acta Crystallogr. C
Struct. Chem. \textbf{70}, 85 (2014).
\end{thebibliography}
\end{document}